\documentstyle [12pt] {article}

\begin{document}
\begin{flushright}
UDHEP-11-95\\
UAAHEP-01-95\\
\end{flushright}
\vspace*{0.5 cm}
\begin{center}
{\bf A Possible Violation of the Equivalence Principle by Neutrinos} \\
\vspace*{1.0 cm}
{\bf A. Halprin} and {\bf C. N. Leung} \\
Department of Physics and Astronomy\\
University of Delaware, Newark, DE 19716 \\
\vspace*{0.5 cm}
and\\
\vspace*{0.5 cm}
{\bf J. Pantaleone}\\
Department of Physics and Astronomy\\
University of Alaska, Anchorage, AK 99508 \\
\vspace*{0.5 cm}
{\bf ABSTRACT} \\
\end{center}

       We consider the effect of a long range, flavor changing
tensor interaction of possible gravitational origin.
Neutrino mixing experiments provide the most sensitive
probe to date for such forces---testing the equivalence
principle at levels below $10^{-20}$.
Here we justify and generalize a formalism for describing such effects.
The constraints from neutrino mixing experiments on gravitationally
induced mixing are calculated.
Our detailed analysis of the atmospheric neutrino data confirms a
remarkable result:  the atmospheric neutrino data
implies the same size force as does the solar neutrino data.
Additional tests of this suggestive result are discussed.

\newpage

\section{ Introduction and Overview}

All experiments sensitive to solar neutrinos have
measured a flux of electron neutrinos smaller than that predicted
by various solar models\cite{BP}, \cite{BU}, \cite{TC}.
It is widely accepted that
the solar neutrino deficit is a result of neutrino
flavor mixing.  The popular mechanism for this mixing is ascribed
to nondegenerate neutrino masses and the inequivalence of the
neutrino weak and mass eigenstates\cite{Pontecorvo}.  A great many
studies have been devoted to deriving constraints on the mixing
parameters: $\Delta m^2$, the neutrino mass-squared difference, and
$\sin^2 2\theta$, the mixing angle, from the solar neutrino data.

In this paper we consider an alternative mechanism of neutrino
mixing in which the flavor changing interaction is assumed to
be gravity\cite{Gasperini}, \cite{HL}.  This assumed violation
of the principle of equivalence
by neutrinos will lead to flavor mixing, even if the neutrinos are
massless.  Consequently, data from neutrino mixing experiments
can be used to test how well the equivalence principle is obeyed
by neutrinos.  As will be seen, current data probe the equivalence
principle to a suprisingly stringent level.

In addition to solar neutrino experiments, recent studies of
atmospheric neutrinos\cite{KIIa}, \cite{Fuk}, \cite{IMB},
\cite{Soudan} also point to the
existence of neutrino mixing.  It is rather intriguing to note that,
in contrast to the conventional mass mixing mechanism, the
gravitationally induced mixing mechanism can account
for the combined solar and atmospheric neutrino data by assuming
{\it only} $\nu_e - \nu_\mu$ mixing.
In the remainder of this paper we shall discuss
the phenomenological consequences of such a flavor dependent
gravitational interaction of the neutrinos.

We begin in Section II with a discussion of how neutrinos
probe for flavor dependence in long range forces.
We start with the  time  delay
in  a  gravitational  field  for  a  massless  particle  and  its
previous applications  to  photons  and  neutrinos.
Next is a discussion of how  this  same  time  delay,  if  flavor
dependent, can produce neutrino oscillations.
Then our formalim is presented; first for massless neutrinos
and then for massive neutrinos.
When nonzero neutrino masses are included into our formalism,
we find a new, physically significant phase parameter.
We conclude Section II with a discussion of the ambiguities of the formalism,
and how some of them can be avoided.
In Section III we
summarize the constraints from accelerator and solar neutrino data;
and we present a new  result  for  the  atmospheric neutrino anomaly.
Also, the parameter regions that planned experiments can probe are estimated.
In Section IV we compare the allowed parameter regions obtained in Section III
and find the remarkable result that atmospheric and solar observations
both select the same, small parameter region for equivalence principle
breaking.
We discuss more stringent tests of this radical idea and mention some possible
outstanding theoretical questions that it generates.

For those readers disinclined to move beyond the orthodoxy of
metric theories of gravitation,  which  preserve  the  equivalence
principle, the results presented here can be used to set the most
restricive bound to date on the validity of the equivalence principle.
They can also be reinterpreted as evidence for the existence of a
new, very long range, flavor changing tensor interaction of
gravitational strength.

\section{Foundations and Formalisms}

\subsection{
Time Delay of Massless Particles in a Gravitational Field}

The most familiar tests of  the  weak  equivalence  principle
(WEP) are experiments of the E\"{o}tv\"{o}s-type\cite{Eotvos}
which measure the gravitational acceleration of macroscopic objects.
Currently, it is found\cite{torsion} that gravity accelerates all
macroscopic objects at the same rate to an accuracy of one part
in $10^{12}$.

A test of the WEP for massless particles stems from Shapiro's
observation that the  transit  time  for  a  photon traversing
a given distance in a gravitational potential, $\phi({\bf r}(t))$,
is dilated by the amount\cite{Shapiro}
\begin{equation}
\Delta t = -(1+\gamma) \int \phi({\bf r}(t))dt ,
\label{Delt}
\end{equation}
where $\gamma$ is a parameter in the parametrized  post-Newtonian
(PPN) formalism \cite{PPN} and the integration is from the time of
emission of the photon to when it  is detected.  In general relativity,
$\gamma = 1$ and radar ranging experiments\cite{radar} have verified
this prediction.  In alternate theories of gravity encompassed
by the PPN formalism and satisfying the WEP, $\gamma$ need not
equal unity, but it must be the same for all particles.
Thus, observation of different $\gamma$-values for two different
particle species would constitute a violation of the WEP.  Limits on such
a violation have been obtained from supernova SN1987A by
comparing the arrival time of photons and neutrinos with the
result\cite{Longo}, \cite{Krauss}
\begin{equation}
|\gamma_\gamma - \gamma_\nu| < {\rm few} \times 10^{-3} .
\label{pn}
\end{equation}
In the same way, comparing neutrinos  with antineutrinos
yields\cite{LoSecco} \cite{PSW}
\begin{equation}
|\gamma_{\nu_e} - \gamma_{\bar{\nu}_e}| < 10^{-6} .
\label{nnbar}
\end{equation}

\subsection{Neutrino Oscillations from Flavor Dependent $\gamma$-Values}

As a  prelude,  we  remind  the  reader  of  the  famous  COW
experiment\cite{COW}.  In this experiment, a beam of neutrons is
passed through a beam splitter and the separated beams are sent
through two spatial regions that are gravitationally inequivalent.
Through this passage the two beams acquire a phase difference
determined by integrals of the  gravitational  potential   through
which they pass.  The split beams are reconstituted and the
diffraction pattern resulting from the gravitationally acquired
phase difference is observed.

Similary, if \(\gamma_\nu\) is flavor dependent, then different
neutrinos will undergo different gravitational time delays
when passing through the same gravitational potential and thereby
acquire different phase shifts.
These phase shifts are observable
owing to the difference in  the  particle  bases  that
diagonalize the weak  and  the  gravitational  interactions.  For
instance, in the case of two neutrino flavors, the weak basis
$(\nu_e, \nu_\mu)$ may be related to the gravitational basis
$(\nu_1,\nu_2)$ as
\begin{equation}
\left( \begin{array}{c}
\nu_e \\
\nu_\mu \end{array}
\right) = \left[ \begin{array}{cc}
\cos \theta_G &  \sin \theta_G \\
- \sin \theta_G & \cos \theta_G \end{array}
\right] \left( \begin{array}{c}
\nu_1 \\
\nu_2 \end{array}
\right),
\label{mix}
\end{equation}
where $\theta_G$ is the mixing angle.
As a consequence, a $\nu_e$
will be able to oscillate into $\nu_\mu$.  (It should be stressed that this
oscillation will take place even if the neutrinos are massless.)  More
specifically, assuming plane wave propagation and a constant potential,
the phase difference due to the time delay acquired by neutrinos of energy
$E$ traversing a distance $l$ is
\begin{equation}
\delta_t = (\gamma_1 - \gamma_2) \phi l E .
\label{delta}
\end{equation}

This phase shift was first described by Gasperini\cite{Gasperini} in terms
of a nonuniversal gravitational red shift.  However,
this result does not include the contribution of phases from the spatial part
of the wavefunction.
In order to combine these two effect in a systematic way,
a more specific model is required.  Such a model was considered
for spinless neutrinos\cite{HL}.  For completeness, we extend
that discussion explicitly to the spin-1/2 case below\cite{IMY}.

In the absence of non-gravitational interactions, the properties of a
spin-1/2 particle in a specified gravitational field, $G_{\alpha\beta}$,
are usually described to first order (linearized therory) by the interaction
Lagrangian density\cite{BD}
\begin{equation}
{\cal L}_{{\rm int}} = i \frac{f}{4} G^{\alpha\beta} [\bar{\psi} \gamma_\alpha
 \partial_\beta \psi - (\partial_\alpha \bar{\psi}) \gamma_\beta \psi] ,
\label{lag}
\end{equation}
where $f = \sqrt{8 \pi G_N}, G_N$ is Newton's constant and the metric
of flat space is $g_{\alpha\beta} = (+1,-1,-1,-1)$.  For our purpose, we
postulate an interaction of the above form but one which allows the neutrinos
$\nu_1$ and $\nu_2$ to couple to gravity with different strengths $f_1$ and
$f_2$.

This breakdown in universality of the gravitational
coupling strength destroys the symmetry that keeps
the graviton massless\cite{Weinberg}. Currently, no
attractive theories have been proposed which can break
the equivalence principle and yet keep the graviton
massless in order to reproduce the experimental result
for the deflection of light\cite{VDV}.  However, such
a theory may be possible\cite{Schwinger}, and it has
been shown that a theory with mixed massless and
massive gravitons can be made consistent with all the
observations\cite{KOS}.  Since the purpose of the
model discussed here is only to put together the
neutrino's temporal and spatial phase shift differences,
and since the graviton mass is not directly relevant to
the neutrino phenomenology, we put aside the formidable
question of the complete consistency of such a theory.

The postulated interaction leads to the equations of motion for the
{\it massless} neutrino fields, $\nu_j$,
\begin{equation}
[(g^{\alpha\beta} + \frac{f_j}{2} G^{\alpha\beta}) \gamma_\alpha
\partial_\beta + \frac{f_j}{4} (\partial_\alpha G^{\alpha\beta})
\gamma_\beta] \nu_j = 0, ~~~~~j = 1, 2, ...
\label{EoM}
\end{equation}
We limit our discussion to situations in which $G^{\alpha\beta}$ varies
very slowly on a scale of order the neutrino Compton wavelength, and
therefore drop the terms involving a derivative of $G^{\alpha\beta}$.  In
this case, we readily find that the $\nu_j$ satisfy a Klein-Gordon equation,
\begin{equation}
(g^{\alpha\beta} + f_j G^{\alpha\beta}) \partial_\alpha \partial_\beta
\nu_j = 0 ,
\label{KG}
\end{equation}
where $O(G^2)$ terms have been dropped for consistency with a linearized
theory.  We assume the gravitational field is determined by a static
macroscopic matter distribution in the harmonic gauge.  (Since general
covariance is broken, the result will in fact be gauge dependent.)  Such a
field is given in terms of the Newtonian potential $\phi$ by\cite{Will122}
\begin{equation}
G_{\alpha\beta} = 2 \phi \delta_{\alpha\beta} / f  ,
\label{hgauge}
\end{equation}
where $\phi(\infty) \rightarrow 0$.


We note from Eqs.(\ref{KG}) and (\ref{hgauge}) that the presence of the
gravitational field modifies the flat space metric in a neutrino species
dependent way such that, for $\nu_j$,
\begin{equation}
g_{\alpha\beta} \rightarrow (g_{\alpha\beta} + \frac{2 f_j}{f} \phi
\delta_{\alpha\beta}).
\label{metric}
\end{equation}
Comparing this with Eq. (1.1) in Ref. \cite{G2} (note that $\phi$ is
defined to be positive in Ref. \cite{G2}), we see that our
approach amounts to the case in which the PPN parameters $\alpha$
and $\gamma$ are identical and the parameters $f_j$ are related to
the PPN parameters by
\begin{equation}
f_j = f \gamma_j.
\label{PPN}
\end{equation}

To illustrate the essential properties of the resulting phase shifts,
we consider the case of constant $\phi$, where we have
the energy-momentum relation
\begin{equation}
E^2 (1 + 2 \gamma_j \phi) = p^2 (1 - 2 \gamma_j \phi) .
\label{E2}
\end{equation}
To first order in $\phi$, the energy eigenvalues of the neutrinos,
$\nu_j$, having the same momentum are given by
\begin{equation}
E_j = (1 - 2 \gamma_j \phi) p .
\label{E}
\end{equation}
For the simple case of two neutrinos, this implies that, after traversing
a distance $l$, the two components,
($\nu_1, \nu_2$), of a state $\nu_e$ will develop a phase difference of
$\delta = 2 (\gamma_1 - \gamma_2) \phi l p$.  If we revert to a
description of  the oscillation  phenomenon utilizing states of equal
energy but differing momenta, the phase difference becomes, to first
order in $\phi$,
\begin{equation}
\delta = 2 (\gamma_1 - \gamma_2) \phi l E ,
\label{del}
\end{equation}
which is twice that obtained from the Shapiro effect alone.

If we compare this phase shift with that obtained in the well known
case of vacuum oscillations induced by a neutrino mass difference,
we find that they are related by the formal connection,
\begin{equation}
\frac{\Delta m^2}{2E} \rightarrow 2E |\phi| \Delta\gamma ,
\label{sub}
\end{equation}
where $\Delta\gamma \equiv \gamma_2 - \gamma_1$.  By analogy,
the $\nu_e$ survival probability after traversing a distance
$l$ is given by
\begin{equation}
P(\nu_e \rightarrow \nu_e) = 1 - \sin^2 2\theta_G \sin^2 \frac
{\pi l}{\lambda}
\label{psurvive}
\end{equation}
where $\lambda$ is the oscillation length which is here given by
\begin{equation}
\lambda = 6.2~{\rm km} \Bigl(\frac{10^{-20}}{|\phi \Delta\gamma |}\Bigr)
\Bigl(\frac{10~{\rm GeV}}{E}\Bigr)
\label{lambda}
\end{equation}
Thus, in sharp constrast to the case of oscillations induced by a
neutrino mass difference where $\lambda$ grows with energy,
gravitationally induced oscillations are characterized by an
oscillation length that diminishes with increasing neutrino
energy.  The two mechanisms may therefore be distinguished
by measuring the neutrino energy spectrum\cite{PHL}, \cite{BKL},
\cite{MN}.

\subsection{Neutrino Oscillations From Both Gravity and Mass Terms.}

In the preceding discussions we have described how neutrino mixing
is generated by breaking the equivalence principle.
It is important to note that this mixing may not preclude
the generation of neutrino mixing from vacuum mass terms---which
is the more commonly discussed source of neutrino mixing.
It is possible that there is flavor dependence in both the neutrino
mass matrix and in the gravitational coupling matrix, simultaneously.
This possibility has been discussed in the literature\cite{G2},
\cite{MN}.  However, these discussions have overlooked an important
point which we explain now.

For simplicity, we shall assume there are only two neutrino flavors.
Choosing to work in the basis which diagonalizes the charged
lepton mass matrix,
the evolution of neutrino flavor in a medium is described by
\begin{eqnarray}
i \frac{d}{dt}
\left( \begin{array}{c} \nu_e \\ \nu_\mu \end{array} \right)
&=&
\left\{
{\Delta m^2 \over 4E}
U_M \left[ \begin{array}{cc}
-1 & 0 \\
0 & 1 \end{array}
\right] U^\dagger_M \right. \nonumber \\
&+& \left.
E |\phi(r)| \Delta\gamma \ \
U_G \left[ \begin{array}{cc}
-1 & 0 \\
0 & 1 \end{array}
\right] U^\dagger_G \right. \\
& + & \left.
{\sqrt{2} \over 2} G_F N_e
\left[ \begin{array}{cc} 1 & 0 \\ 0 & -1 \end{array} \right]
\right\}
\left( \begin{array}{c} \nu_e \\ \nu_\mu \end{array} \right)
\nonumber
\label{osc}
\end{eqnarray}
Here
$\Delta m^2 \equiv m_2^2 - m_1^2$
denotes the difference in neutrino vacuum masses,
$N_e$ is the electron density of the medium, and
$G_F$ is Fermi's constant.
The first term describes the contribution from vacuum masses,
the second term describes the contribution from equivalence
principle breaking,
and the third term describes the contribution from
a background of normal matter\cite{W}.
There are two unitary matrices
which parametrize the mixing,
$U_M$ and $U_G$, and they are generally completely unrelated.
The subscripts $M$ and $G$ denote quantities which come from
the vacuum mass term and from the gravitational mixing, respectively.
A general representation for a unitary matrix is given by
\begin{equation}
U \equiv e^{i \chi}
\left[ \begin{array}{cc} e^{-i \alpha} & 0 \\
0 & e^{i \alpha} \end{array} \right]
\left[ \begin{array}{cc} \cos \theta & \sin \theta \\
-\sin \theta & \cos \theta \end{array} \right]
\left[ \begin{array}{cc} e^{-i \beta} & 0 \\
0 & e^{i \beta} \end{array} \right]
\end{equation}
where $\chi$, $\alpha$ and $\beta$ denote arbitrary phases
and $\theta$ denotes an arbitrary angle.
However, not all of these phases are observable in neutrino
oscillation experiments.

As is well known, when there is no gravitational mixing
(e.g., in a vanishing gravitational field),
all the phases in
$U_M$ may be eliminated by redefinition of the spinors,
e.g., $\nu_e^\prime \equiv {\rm e}^{i 2 \alpha_M} \nu_e $.
Then only $\theta_M$ is observable.
Similarly, when the contribution from the neutrino vacuum masses
is negligble (e.g., in a large gravitational field),
all of the phases in $U_G$ may be eliminated and only
$\theta_G$ is observable.
Thus $\theta_M$ and $\theta_G$ are each independently
observable parameters.
Notice that in general, when neutrino
mixing receives contributions
from both the gravitational and vacuum mass terms,
the $\chi$'s and $\beta$'s can be eliminated
but the $\alpha$'s cannot.
Putting all of the residual phase into the gravitational
term, the flavor evolution may be parametrized as:
\begin{eqnarray}
i \frac{d}{dt}
\left( \begin{array}{c} \nu_e \\ \nu_\mu \end{array} \right)
&=&
\left\{
{\Delta m^2 \over 4E}
 \left[ \begin{array}{cc}
- \cos 2 \theta_M & \sin 2 \theta_M \\
\sin 2 \theta_M & \cos 2 \theta_M \end{array}
\right] \right. \nonumber \\
&+& \left.
E |\phi(r)| \Delta\gamma
 \left[ \begin{array}{cc}
-\cos 2 \theta_G & e^{-i2\alpha} \sin 2 \theta_G \\
e^{i2\alpha} \sin 2 \theta_G & \cos 2 \theta_G \end{array}
\right] \right. \\
&+& \left.
{\sqrt{2} \over 2} G_F N_e
\left[ \begin{array}{cc} 1 & 0 \\ 0 & -1 \end{array} \right]
\right\}
\left( \begin{array}{c} \nu_e \\ \nu_\mu \end{array} \right)
\nonumber
\end{eqnarray}
where $\alpha \equiv ( \alpha_G-\alpha_M )$.
Since this phase cannot be eliminated by redefinition
of the spinors,
it will have experimental consequences.

In the case of constant  backgrounds ($N_e$ and $\phi$),
this equation can be easily solved, yielding the
oscillation probability
\begin{equation}
P(\nu_e , x \rightarrow \nu_\mu , y) =
{ 1 \over 2 } { a^2 \over [ a^2 + b^2 ] }
\left\{ 1 - \cos ( 2 \sqrt{ a^2 + b^2} (y - x) ) \right\},
\label{osc.prob.}
\end{equation}
where $b$ is the diagonal element of the total mixing matrix
\begin{equation}
b \equiv { \Delta m^2 \over 4 E} \cos 2 \theta_M +
E | \phi | \Delta\gamma \cos 2 \theta_G - {\sqrt{2} \over 2} G_F N_e
\end{equation}
$a$ is the magnitude of the off-diagonal element
\begin{equation}
a \equiv \left| { \Delta m^2 \over 4E} \sin 2 \theta_M +
E | \phi | \Delta\gamma \sin 2 \theta_G {\rm e}^{-i 2 \alpha} \right|
\end{equation}
and $x$ and $y$ denote the position of the neutrino source
and detector, respectively.  This solution explicitly shows
that the phase is only observable when both neutrino
vacuum mass mixing and gravitationally induced mixing
are relevant, i.e., when
$\Delta m^2$, $\Delta\gamma$, $\theta_G$ and $\theta_M$ are all
nonzero.

Phases in the vacuum mass mixing matrix may lead to
violations of time reversal (T) symmetry.  Under
this symmetry, neutrino oscillation probabilities
transform as\cite{KPonT}
\begin{equation}
P(\nu_\mu , x \rightarrow \nu_e , y)   \longrightarrow
P(\nu_e, y \rightarrow \nu_\mu , x).
\end{equation}
Unitarity for two neutrino flavors implies that
\begin{equation}
P(\nu_\mu , x \rightarrow \nu_e , y)   =
P(\nu_e , x \rightarrow \nu_\mu , y)
\end{equation}
It follows that a difference in the probabilities
$P(\nu_e , x \rightarrow \nu_\mu , y)$ and
$P(\nu_e, y \rightarrow \nu_\mu , x)$ is a measure of
T violation.  However, note that the presence of a background
medium may lead to an apparent T-violation, even though the
underlying dynamics is T-invariant.  To avoid this one must
consider a background medium which is symmetric about the
midpoint of neutrino propagation, e.g., a constant background
as in Eq. (\ref{osc.prob.}).  It is then clear that the phase
$\alpha$ does not lead to observable T-violations in
two-neutrino oscillations.

In summary, for two neutrinos there are two quantities which
parametrize the strength at which mixing occurs,
$\Delta\gamma$ and $\Delta m^2$, and three quantitites which
parametrize the mixing, $\theta_G$, $\theta_M$ and $\alpha$.
Previous analyses omitted the phase $\alpha$, but in a general
analysis it should be included.
We plan to study the effects of the phase $\alpha$ in more detail
in a future communication.

\subsection{Potential Ambiguities.}

It is necessary to choose a metric in order to confront
experiments.  For any metric choice, breaking the
equivalence principle implies neutrino mixing.  However,
because the particular breaking of the equivalence
principle considered here also breaks general coordinate
invariance, physical results depend on the choice of
metric used.  At present, there are few hints as to
Nature's choice.

In deriving our formalism, we assumed the harmonic metric.
Although this is commonly done in most discussions of
\(K-\bar{K}\) oscillations\cite{kaon}, where one can test
the equivalence principle between a particle and its
antiparticle, the harmonic metric is certainly not the
only possible assumption.  For example, the authors in
Ref. \cite{IMY} assumed the Schwarzschild metric instead.
They have obtained a neutrino oscillation formalism
which only differs from ours by parameter redefinitions.

Some metric choices do, however, lead to observable
consequences.  For instance, the oscillation wavelength
in Eq. (\ref{lambda}) is not
invariant under shifts of $\phi$ by an overall constant.
Thus the expression for $\phi$ that should be used in
calculations is ambiguous, but this ambiguity must be
resolved in order to derive phenomenological constraints.
The most common procedure for fixing a definite value for
$\phi$ is to assume that it vanishes at an infinite
distance from the source.  This insures that as gravity
is "turned off" the results of special relativity are
recovered.  With this assumption, the local gravitational
potentials due to various sources are given in Table 1.
We see that the value of this potential anywhere in our
solar system is dominated by the mass distribution on
scales larger than the galaxy.  However, since the details
of the structure on extra-galactic scales is still a
subject of investigation, the precise value of $\phi$
from these scales is somewhat uncertain.
But it is apparent that, because the dominant $\phi$
comes from scales much larger than the propagation length
in any current neutrino observation,
it is reasonable to ignore the variation
of $\phi$ over the solar system and to take it to be
a constant---the same constant for all local neutrino
experiments.  Since $\Delta \gamma$ and $\phi$ always occur
multiplied together, we can avoid the issue of the
uncertainty in $\phi$ by calculating the constraints on the
product $\phi \Delta \gamma$.  Calculating the constraints
in this manner has several additional advantages.
For the most part, it covers the possibility that $\phi$
does not "turn-off" at large distances,
or that there is some "cosmological" value for $\phi$.
Also, it covers the possibility that $\phi$ is generated
by a long range tensor interaction other than gravity.
Thus, by treating $\phi$ as an unknown constant for our
solar system, we avoid many of the ambiguities inherent to
these calculations.

We briefly mention the interesting possibility that the
coupling between gravity and neutrinos might be spatially
anisotropic.  Such effects have been considered for photons
(see, e.g., \cite{Haugan} and references therein).
For example, the equivalence principle breaking might come
from a coupling of neutrinos to \( \nabla \phi \),
the gradient of the gravitational potential, instead of
to \(\phi\).  This coupling exists in our formalism but
its effects are expected to be drastically suppressed
compared to the leading effects considered here (see the
discussion after Eq. (\ref{EoM})).  This kind of coupling
may also occur in string theories\cite{Polyakov}.
Constraints on such a coupling have been considered
in Ref. \cite{BKL}.  An interesting consequence of this
coupling is that the amount of neutrino mixing depends
on the orientation of the neutrino's momentum
with respect to the gravitational source.

\section{ Experimental Constraints.}

In this section we calculate the experimental limits on
the gravitationally induced mixing ({\it gim}) parameters,
$\Delta \gamma$ and $\theta_G$, from neutrino experiments.

\subsection{Constraints from Rare Muon Decays.}

The anomalous neutrino-gravity interaction breaks the
lepton number symmetry of each family
and will give rise to rare leptonic decay processes.
One might expect a particularly stringent constraint
to arise from the decay $\mu \rightarrow e \gamma$.
Here we consider what limits are placed on the {\it gim}
parameters by this process.

The specific {\it gim} interaction considered in the
preceding section affects only neutrinos.  Consequently
the leading contributions to the decay $\mu \rightarrow
e \gamma$ arise at the one-loop level.  These
contributions turn out not to be renormalizable.
However, it might be argued that the equivalence
principle is broken at a scale that still preserves
the weak isospin symmetry.  Since gravitation is not
(yet) describable as a renormalizable gauge theory,
the argument is suspect.  Nevertheless, we shall examine
this scenario, since it will generate the rare muon
decay at tree level and presumably provide the most
restrictive calculable constraint on the equivalence
principle violating parameters obtainable from this
process.  While the Standard Model symmetry allows an
additional equivalence principle violating term for the
right-handed charged leptons, we shall ignore this
complication.  We simply replace the (left-handed)
neutrino fields in their anomalous gravitation
interactions by the corresponding left-handed family
isospin doublets and make the derivatives appearing
there SU(2) gauge covariant in the usual fashion.  This
leads to three tree-level Feynman diagrams which yield
a decay rate of order magnitude
\begin{equation}
\Gamma \sim \alpha m_{\mu} (\Delta\gamma \phi \sin 2\theta_G)^2,
\label{width}
\end{equation}
where $\alpha$ is the fine structure constant and
$m_{\mu}$ is the muon mass.  Taking the large but
somewhat uncertain value of $|\phi| = 3 \times 10^{-5}$,
we estimate the branching ratio for $\mu \rightarrow
e \gamma$ to be
\begin{equation}
B(\mu \rightarrow e \gamma) \sim 2.3 \times 10^6 (\Delta\gamma
\sin 2 \theta_G)^2,
\end{equation}
Since the current experimental limit\cite{PDG} is
$B(\mu \rightarrow e \gamma) < 5 \times 10^{-11}$,
this leads to the constraint
\begin{equation}
|\Delta\gamma \sin 2\theta_G| < 5 \times 10^{-9} .
\end{equation}
Except for very small mixing angle, this constraint on
$\Delta\gamma$ is inferior to the constraints obtained
below from solar, atmospheric and accelarator neutrino
experiments. \\

\subsection{Constraints from Accelerator Experiments.}

There exist numerous constraints on
neutrino mixing from experiments using neutrinos
produced by accelerators.  Because accelerators
produce a neutrino flux which is mostly muon-neutrinos
with a much smaller electron-neutrino component,
these experiments mostly probe $\nu_\mu-\nu_x$ mixing
with a reduced sensitivity to $\nu_e-\nu_x$ mixing.
Accelerator experiments typically involve neutrinos
with energies ranging from a GeV to hundreds of GeV,
and propagation lengths up to a kilometer.  Since the
neutrino cross section increases with energy, these
experiments can achieve large event rates and so can
probe for relatively small values of the mixing angle.
Most importantly for us, the high neutrino energies in
these experiments make them powerful tools for testing
the equivalence principle because the gravitationally
induced mixing `turns on' with increasing energy.

It is difficult to utilize the published analyses to
obtain the exclusion region for $|\phi \Delta\gamma|$
and $\sin^2 2\theta_G$.  This is because the energy
dependence of {\it gim} is so radically different
from the case of mass mixing that the analyses must be
completely redone.  However, many important details
such as the energy dependence of the observed and
expected event rates are seldom published.
We are thus forced to a less than optimal
estimate of the exclusion region.  Here we calculate
the exclusion region for acclerator experiments
in the long wavelength and the short wavelength limits.
These lines are then extended to intersect in order to
estimate the full exclusion region.
The constraint in the short wavelength limit is obtained
by taking the average probability at minimum sensitivity
to be the same for both mixing mechanisms.
The constraint in the long wavelength limit is obtained by
using the minimum $\Delta m^2$ from the published analysis
to calculate the typical energy scale of the experiment and,
with the average length given in the published reference,
we compute the minimum value for $|\phi \Delta \gamma|$.
In this admittedly crude fashion, we obtain the excluded
regions shown by the straight-line contours in Fig. (1).
We estimate that these contours are accurate up to
factors of 3 in $|\phi \Delta \gamma|$.  We have selected
experiments\cite{accelerator} having the largest values of
$E \cdot l$, for they provide the most stringent limits.

Recently, the LSND accelerator neutrino
experiment has reported some candidate events
that may\cite{LSND}, or may not\cite{Hill},
indicate ${\bar \nu}_\mu-{\bar \nu}_e$ neutrino
oscillations.  The favored parameter region lies
above $|\phi \Delta \gamma| > 5 \times 10^{-18}$
and so large mixing angles are excluded by other,
higher energy, accelerator neutrino
experiments\cite{Mann}.  There may be a very small
parameter region, allowed by all accelerator experimental
results, at $\sin^2 2 \theta \approx 10^{-3}$
and $|\phi \Delta \gamma| \approx 2 \times 10^{-16}$.
In light of the preliminary nature of the LSND
experimental results, we shall not pursue this
analysis further.

There are several serious proposals for a new
generation of accelerator experiments with
neutrino propagation lengths of several
hundreds kilometers\cite{MINOS} \cite{otherLBL}.
For example, the MINOS experiment would send a
neutrino beam from the Fermilab accelerator
to the Soudan-2 detector located 730 kilometers
away.  The lower, curved contours in Fig. (1) estimate
the {\it gim} parameter region that can be probed by
this experiment. Note that matter effects start
to become important at these longer distances and
cause a difference in the neutrino and antineutrino
oscillation probabilities, as shown by the solid
and dotted curves in Fig. (1).

\subsection{Constraints from Reactor Experiments.}

Numerous constraints on neutrino mixing have been
obtained from experiments using neutrinos produced
by nuclear reactors.  Nuclear reactions create
neutrinos with energies ranging from a fraction of
an MeV to order 10 MeV, and these have been detected
hundreds of meters from the reactors.  While these
relatively low energy neutrinos provide stringent
constraints on mass mixing, they provide only very
weak constraints on {\it gim} because of the
contrasting energy dependence of the two mechanisms.
Specifically, reactor experiments only probe values
of $|\phi \Delta \gamma| > 1 \times 10^{-16}$.  For
the {\it gim} mechanism, reactor constraints are
completely surpassed by those from accelerator
experiments, so we shall not discuss them further.

\subsection{Solar Neutrino Constraints.}

The nuclear reactions which power the sun also produce
neutrinos.  These neutrinos have been observed here on
Earth by several experiments\cite{CL} \cite{KII}
\cite{GALLEX} \cite{SAGE}.  These experiments study
different ranges of the neutrino spectrum, which extends
up to 14 MeV and is a superposition of several components
(see, e.g., \cite{BP}).  All of these measurements find
far fewer neutrinos than expected (see Table 2).  This
is commonly interpreted as evidence for neutrino mixing.
This is because both the production and detection of solar
neutrinos primarily involve only the $\nu_e$ flavor,
so any mixing will reduce the observed flux.
Here we shall assume that all of the neutrino mixing
comes from the {\it gim} mechanism,
and derive the implications of solar measurements for {\it gim}
parameters.

The oscillation probabilities are calculated analytically.
Here, in accordance with the discussion in
Section 2.4, we shall take $\phi$ to be a constant
(in our earlier analysis\cite{PHL} we used the $\phi$ generated
by the Sun).
{\it Gim} effects start being important for solar neutrinos
when $ |\phi \Delta \gamma | > 2 \times 10^{-25}$.  This is
when a 10 MeV neutrino undergoes half of an oscillation
in its propagation from the Sun to the Earth.
For larger values of $|\phi \Delta \gamma |$ the oscillation wavelength,
Eq. (\ref{lambda}), is smaller, and when the wavelength is less than the
size of the Sun, the effects of the background matter become important.
The analysis of matter effects\cite{W}
parallels the well known case for mass mixing\cite{MS}
(for a review, see, e.g., \cite{KP}),
with the substitutions of Eq. (\ref{sub}) (for constant potential) and
$\theta \rightarrow \theta_G$.  The condition for a resonance to
occur is now given by
\begin{equation}
\sqrt{2} G_F N_e = 2 E |\phi| \Delta\gamma \cos 2 \theta_G ,
\label{res}
\end{equation}
where \(N_e\) is the electron density in the sun.
Note that when matter effects are relevant,
the sign of $\Delta \gamma$ is important.
For $\Delta \gamma > 0$ the neutrinos can go through
a resonance, and if the transition is adiabatic,
large reductions in the neutrino flux occur.
The calculated oscillation probability, for $\Delta \gamma > 0$,
is shown in Fig. (2) as a function of neutrino energy, $E$,
times $ |\phi| \Delta \gamma$.
Scanning across Fig. (2) from low to high energies, we see that
first the long wavelength oscillations occur,
then the nonadiabatic side of the MSW well, and then the
adiabatic side.  This order is a reversal from that of
the mass mixing mechanism, which provides a means to
experimentally distinguish the two mechanisms.

The most recent solar neutrino data and solar model
results are used in our calculation of the constraints
on {\it gim} parameters.  In addition to the results
given in Table 2, the Kamiokande group has looked at
how their data depends on energy\cite{KII}.  These
data can be used to constrain mixing, since mixing
effects can be energy dependent (see Fig. (2)).
Much of the neutrino energy dependence is lost in the
neutrino-electron scattering process, which occurs in
the Kamiokande detector, but enough remains to be
important.  We have folded our oscillation probabilities
into the neutrino spectrum and then integrated this
times the cross section over energy to get the predicted
result for each Kamiokande energy bin, and for each of
the other experiments.  The total chi-squared is
calculated between the predicted and measured results
and this is used to find the allowed {\it gim} parameter
regions.  Only the experimental errors were included,
since these are estimated to be the largest.  We first
calculated the allowed parameter region in the mass
mixing mechanism.  We then compared these calculations
with those of others (e.g., \cite{Langacker}) and found
that our 90\% contours were in good agreement with
theirs.  We used the same value of chi-squared to
calculate the corresponding 90\% contours in the {\it gim}
mechanism.  This insures that the mass mixing mechanism
and the {\it gim} mechanism are treated equally in
fitting the same data.

The results of our analysis are given in Fig. (3).
There are plotted the allowed parameter regions at 90\%
and 99\% confidence levels, assuming $\Delta \gamma > 0$.
For $\Delta \gamma < 0$ there is no allowed parameter
region because then matter effects suppress mixing.
The updated data, and the use of a constant $|\phi|$,
do not significantly alter the allowed parameter
regions from those found in our earlier
analysis\cite{PHL}, which were independently
confirmed\cite{BKL}.  There are two allowed parameter
regions: one at large mixings, where the MSW effect
suppresses mostly the intermediate and high energy
neutrinos, and another at small mixings, where the
MSW effect suppresses mostly the intermediate energy
neutrinos.  These two parameter regions are analogous
to the two MSW regions found for the mass mixing
mechanism (see, e.g., \cite{Langacker}).

For the mass mixing mechanism there is also a long
wavelength, vacuum oscillation solution (see, e.g.,
\cite{long}).  For the {\it gim} mechanism, long
wavelenth effects are relevant for large vacuum
mixings and $|\phi \Delta \gamma | \approx 2 \times
10^{-25}$.  Around this point only the highest energy
solar neutrinos are suppressed (see Fig. (2)).  However,
the data require some suppression at low neutrino
energies so there is no allowed {\it gim} parameter region near
this value\cite{PHL}.

There are several new solar neutrino experiments
which will test these explanations of the data.
The SNO\cite{SNO} experiment will definitively test
all neutrino mixing explanations of the solar
neutrino deficit by performing a flavor independent
measurement of the solar neutrino flux.  The SNO and
Super-Kamiokande\cite{SK} experiments will be able
to measure the energy dependence of the high energy
solar neutrino flux.  The large mixing solution would
not give an energy dependence observable in these
experiments, but the small mixing solution would.
Both of these experiments can also look for day-night
variations from a resonance with matter in the Earth.
These variations are large for these experiments when
the mixing angle is large and when
$0.75 < |\phi| \Delta \gamma / 10^{-20} < 2.5 $.
Thus, if the experiments are sensitve to small day-night
variations, part of the large mixing solution may be
probed by this method.  In addition, the BOREXINO
experiment\cite{BOREXINO} will measure the flux of
Be$^7$ neutrinos with high statistics.  This flux is
almost monoenergetic so long-wavelength oscillations
are important for a wide range of parameters.  By looking
for temporal variations on scales from a few days to a
few months, correlated with changes in the Earth-Sun
distance, this experiment can probe the parameter region
\begin{equation}
3 \times 10^{-4} < {|\phi \Delta \gamma | \over 10^{-20}} < 0.3
\end{equation}
at large values of the mixing angles\cite{PP}.
Thus BOREXINO can probe all of the large mixing angle
solution.  These new probes using neutral current flux
measurements, spectral distortions and temporal variations
are especially important because they are independent of
uncertainties in the solar model.

\subsection{Connection with Supernova Dynamics.}

In a stellar collapse, neutrinos are created in large
numbers in the hot, high density core, and then
diffuse out to lower matter densities where they
then freely pass through the outer layers of the star.
If the neutrinos were to go through a resonant flavor
conversion in the region near the supernova's core,
this would have important, observable effects.
Present models of supernovae suggest that this would
increase the size of the explosion by about 50\%,
and it would also block the production of r-process
nuclides\cite{Qian}.  The first effect would be a welcome
one, since historically theoretical models have
fallen short of the observed energy.  However, the second
effect would eliminate one of the most promising sites for
r-process nucleosynthesis (see, e.g., \cite{Woosley} and
references therein).  Our knowledge of supernova dynamics
is presently too sketchy to allow reliable constraints to
be placed on any neutrino mixing parameters.  Here we
estimate the {\it gim} parameter region that is relevant
for supernova dynamics.

Qualitatively, supernova neutrinos propagate from high
to low densities just as do solar neutrinos.  However,
calculating the probability that a supernova neutrino's
flavor survives a resonant transition is technically
more difficult because the large neutrino background in
a supernova makes the flavor evolution
nonlinear\cite{nonlinear}.  A recent numerical study
has found that this effect substantially reduces the
relevant parameter region\cite{Sigl}.  Using the results
of this study, we estimate that {\it gim} parameters in
the range
\begin{equation}
\Delta \gamma > 4 \times 10^{-14} \ \ ,
\ \ \ \ \sin^2 2 \theta > 10^{-3}
\label{sup}
\end{equation}
would have substantial effects on supernova dynamics.
In this estimate we took the gravitational potential at
the nucleosynthesis region of a supernova to be $10^{-1}$.
Because this value is larger than those in Table 1,
we did not take $\phi$ to be a constant determined by
external sources in deriving Eq. (\ref{sup}) (see the
discussion in Sec. 2.4).  To compare the parameter region
relevant for supernovae with those obtained in local
neutrino observations, we must multiply the above limit on
$\Delta \gamma$ by the value of $\phi$ relevant to the
solar system.  Taking this to be $3 \times 10^{-5}$, the
supernova parameter region lies just above the parameter
region probed by the solar neutrino measurements.

\subsection{Atmospheric Neutrino Constraints.}

Recent experimental studies\cite{KIIa} \cite{Fuk} \cite{IMB}
\cite{Soudan} of neutrinos produced in the Earth's
atmosphere have found evidence for neutrino mixing.
Atmospheric neutrinos have properties very different
from solar neutrinos.  Atmospheric neutrinos are more
energetic, with energies ranging from fractions of
a GeV to several thousand GeV.  Furthermore, their
propagation length is shorter; it varies from 20 to
13~000 km.  In the mass mixing mechanism, the
oscillation phase depends on $length/energy$, so this
quantity is much smaller for atmospheric neutrinos than
it is for solar neutrinos.  Consequently the mass mixing
mechanism requires mixings of different neutrino pairs
to fit the combined solar and atmospheric neutrino data
(see, e.g., \cite{as3}).  In contrast the oscillation
phase in the {\it gim} mechanism depends on
$length \cdot energy$, which has the same order of
magnitude for both the solar and the atmospheric
neutrinos\cite{PHL}.  This offers the suggestive
possibility that the solar and atmospheric neutrino data
can be simultaneously explained with the same neutrino
mixing parameters.  This expectation is indeed confirmed
by the results of our analysis discussed below.

The most precise measurements of the atmospheric neutrino
flux have been made with the large water Cherenkov detectors
Kamiokande\cite{KIIa} \cite{Fuk} and IMB \cite{IMB}.  These
detectors each have exposures which are more than 5 times
larger than that of any other detector.  Here we confine our
analysis to their results.  Many quantites used in the
calculation of neutrino event rates have large uncertainties.
For example, the prediction for the absolute atmospheric
neutrino event rates has an estimated uncertainty of about 30\%.
Thus, one must be careful about which experimental results are
used to constrain neutrino mixing.  We do not use the experimental
results of FREJUS\cite{Frejus} because of their smaller statistics,
and because their results and those in Ref. \cite{IMBbad}
depend on the neutrino-rock cross section for which sizeable
corrections have been calculated recently\cite{LLS}.
The predictions for the relative rate of events starting in the
detector is believed to be much more reliable, with an
estimated uncertainty of only about 5\%.  Neutrino mixing
could change the relative number of $\nu_\mu$ to $\nu_{e}$
in the flux from the expected value of approximately 2.  Thus, the
relative rate can be used as an indicator of neutrino mixing effects.

Kamiokande and IMB have both measured the ratio of $\nu_\mu$ to $\nu_{e}$
for events fully contained in their detector.
Dividing this number by the ratio predicted by a Monte Carlo analysis gives
\begin{equation}
R \equiv { {N_\mu^{exp} / N_{\rm e}^{exp} } \over
{N_\mu^{MC} / N_{\rm e}^{MC} }  }  \  \ .
\end{equation}
In the "sub-Gev" energy range (0.3 - 1.3 GeV for Kamiokande, and
0.3 - 1.5 GeV for IMB),
this ratio is found to be
\begin{eqnarray}
R = 0.60 \pm 0.05 \pm 0.05 \   \ ({\rm Kam}) \nonumber \\
R = 0.54 \pm 0.05 \pm 0.12 \   \ ({\rm IMB})
\end{eqnarray}
Kamiokande has also measured this ratio for higher energy
(1.3 GeV to around 10 GeV) neutrinos, which involve fully
contained as well as partially contained events.  They
find this "multi-Gev" ratio to be
\begin{equation}
R = 0.57 \pm 0.07 \pm 0.07 \   \ ({\rm Kam})
\end{equation}
These consistent departures from 1.0 suggest neutrino mixing.

In addition, the Kamiokande group has studied how their data
depends on the zenith angle.  As the angle between the neutrino
and the zenith varies from 0 to $\pi$, the neutrino propagation
length varies from 20 to 13~000 km.
Both mass mixing and {\it gim} predict that
mixing effects should appear at "long" distances,
so an unusual angular dependence could indicate neutrino mixing.
The cross section in the "sub-Gev" energy range is
relatively isotropic, so no evidence for an angular dependence
is expected or found in those data.  In the "multi-GeV"
energy region, the cross section is more directional, with the
average angular spread between the neutrino and its associated
charged lepton being $\Theta_{rms} \approx 15^\circ-20^\circ$.
There Kamiokande has found an angular dependence in their data,
with longer path lengths associated with smaller R values\cite{Fuk}.
This also suggests neutrino mixing.

Neutrino mixing modifies the observed event rates as given by
\begin{eqnarray}
N_e & = & N_e^{\rm MC} [ P(\nu_e \rightarrow \nu_e) +
r P(\nu_\mu \rightarrow \nu_e) ] \nonumber \\
N_\mu & = & N_\mu^{\rm MC} [ P(\nu_e \rightarrow \nu_\mu)/r +
 P(\nu_\mu \rightarrow \nu_\mu) ]
\end{eqnarray}
Here $N_\alpha$ denotes the number of charged leptons of type $\alpha$
observed in a particular bin, and the MC superscript denotes the
number expected without oscillations from a Monte Carlo calculation.
The ratio of the $\nu_\mu$ flux to the
$\nu_e$ flux is denoted by $r$ and is
approximately 2.1 \cite{HKM}.
The P's are the oscillation
probablities averaged over the energy and length distributions
relevant for a particular bin.
In the calculations performed by the experimental groups,
the data is broken up into an array of bins depending on energy and
zenith angle.  However these full data arrays are not published.
For our calculations, all of the sub-GeV data lies in just
one bin, while the multi-GeV data is divided into 5 zenith angle bins.

The oscillation probabilities are calculated analytically.
Here, as for the earthbound accelerator neutrino experiments
and the solar neutrino measurements, $\phi$ is taken to be
a constant (see discussion in Section 2.4).  For $\nu_\mu-\nu_e$
oscillations matter effects are accounted for by using a two
density (core and mantle) model for the earth,
assuming $\Delta \gamma > 0$,
and taking $3/7$ of the data to be due to antineutrinos.
The oscillation probabilities are averaged over
neutrino energy distributions
as given in references \cite{Fuk} and \cite{Gaisser}.
The average over length distributions is calculated as
\begin{equation}
\langle P_\nu \rangle~\propto~\int d \Omega P_\nu(\theta)
I_\nu (\theta) \int_{bin} d \Omega' P(\theta,\phi;\theta',\phi')
\end{equation}
where $\theta$ is the zenith angle, $P_\nu$ denotes the
neutrino oscillation probability, $I_\nu$ is the neutrino
flux intensity at the detector without mixing, and
$P(\theta,\phi;\theta',\phi')$ denotes the probability
of a neutrino with angular coordinates $\theta,\phi$ giving
rise to a charged lepton with angular coordinates
$\theta',\phi'$.  For the sub-GeV data the last quantity is
not necessary, but for the multi-GeV data we take it to be
a gaussian with an rms spread of 20 degrees\cite{Fuk}.
For $I_\nu$, we assume that at their production point
in the atmosphere the neutrino fluxes are independent of
angle and energy.  This is a good approximation for the
multi-GeV energy range.  With this assumption, $d\Omega I_\nu$
at the detector is proportional to the particularly
simple form of $dL / L$ where L is the neutrino
propagation length.

We treat the "sub-GeV" and "multi-GeV"
data sets independently, because the relative errors
(e.g., in the cross section, etc.)
between these two energy ranges have not been studied.
For each data set we calculate the $\chi^2$ between the
predicted and observed event rates\cite{FL}.
In addition to statistical errors, we include a
$30\%$ error for the overall flux normalization.
We do not explicitly include other, much smaller, errors in
our $\chi^2$'s (e.g., those from particle misidentification,
the Monte Carlo calculations, etc).
Instead we first calculated the allowed parameter
region in the mass mixing mechanism.
We then compared these calculations to those of the
Kamiokande experimental group\cite{Fuk}, and found that
our model gave 90\% contours that were in good agreement
with theirs for particular, reasonable choices of $\chi^2$.
We then used these values of $\chi^2$ to calculate the
corresponding $90\%$ contours in the {\it gim} mechanism.
This procedure was adopted for several reasons:
it tests our model and insures that it
is reasonable and accurate, it includes as many
unknown experimental effects as possible into our calculations,
and it insures that the mass mixing mechanism
and the {\it gim} mechanism are treated equally in
fitting the same data.

The regions of allowed {\it gim} parameters are shown in
Figs. (4).  Fig. (4a) is for $\nu_e-\nu_\mu$ mixing and
Fig. (4b) is for $\nu_\mu-\nu_\tau$ mixing with
$\Delta \gamma > 0$ (for $\Delta \gamma < 0$ the contours
are almost identical).  The parameter regions allowed by
the sub-GeV and multi-GeV data sets lie to the right of
the dashed and solid contours, respectively.  The sub-GeV
measurements are insensitive to any angular dependence in
the flux, so they allow arbitrarily large values of
$| \phi | \Delta \gamma$ where the oscillations completely
average out.  However, the multi-GeV data contains some
angular dependence so it extends only over a finite range
of $|\phi| \Delta \gamma$.  Because higher energy neutrino
experiments are more sensitive to {\it gim} effects (see Eq.
(\ref{psurvive}) and (\ref{lambda})), the bottom of the
multi-GeV contours lies below the bottom of the sub-GeV contours.
The overlapping region is where both data sets can be simultaneously
explained.  For $\nu_e-\nu_\mu$ mixing, Fig. (4a), the overlapping
region is relatively large and contains the best fit points for
both data sets.  Thus the sub-GeV and multi-GeV atmospheric
neutrino data can be consistently explained by the {\it gim}
mechanism.

There are several new atmospheric neutrino measurements
which will test this explanation of the data.
Super-Kamiokande\cite{SK} will increase the precision
of the existing Kamiokande measurement.
Soudan-2 \cite{Soudan} will measure the flux in a detector
that is not a water Cherenkov type experiment.
In addition, the next generation of neutrino
telescopes\cite{telescopes},
DUMAND, AMANDA, NESTOR and Baikal,
may be able to accurately measure the atmospheric neutrino flux
at energies which are an order of magnitude larger.
While the mass mixing mechanism turns off at these higher energies,
the {\it gim}  explanation of the data does not.
As these experiments measure the angular dependence of the flux,
they will probe a parameter region an order of magnitude
below that probed by the current atmospheric neutrino experiments.

\section{ Discussion and Conclusions.}

We have reviewed all of the present experimental constraints
on neutrino mixing for their implication on equivalence
principle breaking parameters.  The most stringent constraints
are from: present accelerator neutrino experiments, which are
sensitive down to $|\phi| \Delta \gamma \approx 10^{-21}$;
solar neutrino experiments, which are sensitive
down to $|\phi| \Delta \gamma \approx 10^{-22}$; and
atmospheric neutrino experiments, which are sensitive
down to $|\phi| \Delta \gamma \approx 10^{-23}$.
Currently, the latter two types of measurements actually
indicate nonzero neutrino mixing.  Assuming that this
mixing arises solely from a violation of the equivalence
principle, we have analyzed these measurements to find the
allowed values of the mixing parameters.  A comparison of
Figs. (3) and (4a) shows that the solar and atmospheric
neutrino measurements can be simultaneously explained
by $\nu_e-\nu_\mu$ mixing with
$0.6 < \sin^2 2 \theta_G < 0.9$ and
$2 \times 10^{-22} < |\phi \Delta\gamma| < 2 \times 10^{-21}$.
This is an extremely remarkable result since the two types of
events are very different in their characteristic
neutrino energies and propagation lengths.

Because gravitationlly induced mixing has an energy
dependence which is the inverse of the mass mixing
mechanism, the two mechanisms give quite different
predictions.  Accelerator neutrino experiments offer
a controlled, independent test of the {\it gim}
solution.  These experiments already rule out the
upper half of the parameter region allowed by the
solar and atmospheric neutrino data (see Fig. (1)).
Next generation long-baseline experiments such as
MINOS promise to extend these limits by over 3 orders
of magnitude.  This solution will also be tested as
the next generation of solar and atmospheric neutrino
experiments start up in the near future.  In
particular, the BOREXINO solar neutrino experiment
can use long-wavelength vacuum oscillations to
directly probe the allowed parameter region,
independent of solar model uncertainties.

The test of the equivalence principle discussed in this
paper is quite similar to the more familiar tests such as
E\"{o}tv\"{o}s-type experiments, Shapiro's time dilation
experiments, and others.  However, there are theoretical
difficulties attendant with a violation of the weak
equivalence principle.  For instance, it naievely suggests
the presence of a graviton mass on which there are stringent
experimental constraints from data on the bending of light.
The importance of this difficulty is not clear since
a consistent theory of quantum gravity has proved elusive.
Of course, our result certainly does not
compel a violation of the equivalence principle.
An alternate interpretation might be the existence of a very
long range tensor field, in addition to the gravitational
tensor, that couples to electron- and muon-neutrinos in the
manner described here.  To distinguish this alternative would
require positive indications with particles other than
neutrinos---an experimentally challenging task.  Neutrinos
are uniquely suited for testing the equivalence principle
because they are subject to only the two weakest known forces,
the weak interactions and gravity.

Experiments currently under construction will decide if the
equivalence principle is violated, as the current evidence
suggests.  If this parametrization ultimately proves
inadequate by more refined experimental results, this type
of analysis will reinforce the concept of a geometric
theory of gravity.  If this class of mechanisms provides
the proper parameterization of the data, then perhaps we
have at hand the Balmer formula for neutrino oscillations.

\newpage

\noindent {\bf Acknowledgements}\\

A.H. would like to thank Sandip Pakvasa and Xerxes Tata
for useful conversations. He would also like to thank the
Center for Theoretical Physics of Seoul National University
and the Laboratoire de Physique Theorique et Houtes
Energies of the University of Paris XI for hospitality  during
periods of manuscript preparation. Two of the  authors
(A.H.  and J.P.) are also indebted to the Nuclear Theory
Institute of  the University of Washngton for hospitality and
stimulating conversation during its 1994 Solar Neutrino
Workshop.  C.N.L. would like to thank the International School
for Advanced Studies, especially S. T. Petcov, for their
hospitality during the initial writing of this manuscript.
He would also like to thank S. T. Petcov, A. Yu Smirnov and
P. I. Krastev for useful discussions.  This work was supported
in part by the United States Department of Energy under Grant No.
DE-FG02-84ER40163.

\raggedbottom

\raggedbottom
\newpage

%
%

\noindent{{\bf Table 1.} Values of the gravitational
potential $ |\phi| \equiv G_N M /r$
at various positions from various sources\cite{MTW}.
The Sun-Sun entry is the largest value of $|\phi|$
in the Sun due to the Sun.
The details of structure on supercluster scales are
not well measured at present, so there is a sizeable
uncertainty in the last entry\cite{Kenyon}.} \\

\begin{center}
\begin{tabular}{| l | l | l |} \hline
Position   & Source  & $|\phi|$ \\ \hline
Earth      & Earth   & $6 \times 10^{-10}$ \\
Earth      & Sun     & $1 \times 10^{-8}$ \\
Solar sys. & Galaxy  & $6 \times 10^{-7}$ \\
Solar sys. & Virgo cluster & $1 \times 10^{-6}$ \\
Sun        & Sun     & $7 \times 10^{-6}$\\
Solar sys. & Great Attractor & $3 \times 10^{-5}$ \\
\hline
\end{tabular}
\end{center}

\newpage

\noindent{{\bf Table 2.}  Results of the Homestake\cite{CL},
Kamiokande-II\cite{KII}, and combined SAGE\cite{SAGE}
and GALLEX\cite{GALLEX} solar neutrino experiments.}

\begin{center}
\begin{tabular}{| l | l | l | l | l |} \hline
Experiment & Process & E\(_{threshold}\) & Rate (SNU)
& Theory (SNU) \\ \hline
Homestake & \(\nu_e + ^{37}\)Cl\( \rightarrow e + ^{37}\)Ar
   & 0.81 MeV  & 2.5 \(\pm\) 0.3 & 8.0 $\pm$ 1.0 \\
Kamiokande-II & \(\nu + e \rightarrow \nu + e \)
   & 7.5 MeV  & 0.50 \(\pm\) 0.07$^*$ & 1.0 $\pm$ 0.15$^*$ \\
GALLEX+SAGE& \(\nu_e + ^{71}\)Ga\( \rightarrow e + ^{71}\)Ge
   & 0.24 MeV  & 78  \(\pm\) 10 & 137 $\pm$ 8   \\ \hline
\end{tabular}
\end{center}
\vspace*{0.3cm}

\noindent{$^*$The Kamiokande flux is not given in SNU, but
as a fraction of the standard solar model\cite{BP}
prediction.}

\raggedbottom

\newpage

\begin{center}
{\bf FIGURE CAPTION} \\
\end{center}
\vspace*{0.6cm}

\noindent {\bf Fig. (1).}  Constraints on {\it gim} parameters
          from present and proposed accelerator neutrino
          experiments.  To the right of the straight-line
          contours lie the regions of $|\phi \Delta \gamma|$ and
	  $\sin^2 2\theta_G$ that are excluded by current
          accelerator experiments (see Sec. 3.2 and Ref. \cite{accelerator}).
	  To the the right of the curved contours lie the regions
	  that may be probed by MINOS, a planned long-baseline
	  accelerator experiment, assuming a 10\% sensitivity for
	  a disappearance experiment and $\Delta \gamma > 0$.
	  The outer, solid curve is for $\nu_\mu$ while
	  the inner, dotted curve is for ${\bar \nu}_\mu$. \\

\noindent {\bf Fig. (2).}  Plot of $P(\nu_e \rightarrow \nu_e )$
	  as a function of $E |\phi| \Delta \gamma$
	  for $\sin^2 2 \theta_G = 0.4$.  The
	  oscillation probability has been averaged over the $^8$B
	  neutrino production region of the Sun.  \\

\noindent {\bf Fig. (3).}  $\chi^2$ plot showing regions of
	  $|\phi| \Delta \gamma$ and $\sin^2 2\theta_G$ allowed by
	  the solar neutrino data in Table 2 at 90\% (solid
	  lines) and 99\% (dashed lines) confidence level,
	  assuming two neutrino $\nu_e -\nu_x$ mixing. \\

\noindent {\bf Fig. (4).}  Plot of $|{\phi}| \Delta \gamma$
	  versus $\sin^2 2\theta_G$ showing parameter regions allowed
	  by the atmospheric neutrino data at 90\% confidence level.
	  Two flavor neutrino oscillations are assumed with
	  $\nu_e-\nu_\mu$ oscillations in Fig. (4a) and $\nu_\mu-\nu_\tau$
	  oscillations in Fig. (4b).  The region allowed by the sub-GeV
	  data lies to the right of the dashed contours, and the region
	  allowed by the multi-GeV data lies to the right of the solid
	  contours.  The best-fit points are also shown by a triangle
	  for the sub-GeV data and a cross for the multi-GeV data. \\

\end{document}